\documentclass[12pt,prd,tightenlines,nofootinbib,showpacsshowkeys]{revtex4}
\newcommand{\be}{\begin{equation}}
\newcommand{\ee}{\end{equation}}
\newcommand{\bga}{\begin{equation}\begin{gathered}}
\newcommand{\ega}{\end{gathered}\end{equation}}
\newcommand{\bgas}{\begin{equation}\begin{gathered}}
\newcommand{\egas}{\end{gathered}\end{equation}}

\usepackage{graphics}
\usepackage{rotating}
\usepackage{epsfig}
\usepackage{amsmath}
\usepackage{amsfonts}

\begin{document}
\title{\begin{flushright}{\rm\normalsize SSU-HEP-14/06\\[3mm]}\end{flushright}
Hyperfine structure of $S$-states in muonic deuterium}
\author{\firstname{R.~N.} \surname{Faustov}}
\affiliation{Dorodnicyn Computing Centre, Russian Academy of Science, Vavilov Str. 40, 119991, Moscow, Russia}
\author{\firstname{A.~P.} \surname{Martynenko}}
\affiliation{Samara State University, Pavlov Str. 1, 443011, Samara, Russia}
\affiliation{Samara State Aerospace University named after S.P. Korolyov, Moskovskoye Shosse 34, 443086,
Samara, Russia}
\author{\firstname{G.~A.} \surname{Martynenko}}
\author{\firstname{V.~V.} \surname{Sorokin}}
\affiliation{Samara State University, Pavlov Str. 1, 443011, Samara, Russia }

\begin{abstract}
On the basis of quasipotential method in quantum electrodynamics we calculate corrections
of order $\alpha^5$ and $\alpha^6$ to hyperfine structure of $S$-wave energy levels
of muonic deuterium. Relativistic corrections, effects of vacuum polarization in first, second
and third orders of perturbation theory, nuclear structure and recoil corrections are taken into account.
The obtained numerical values of hyperfine splitting $\Delta E^{hfs}(1S)=50.2814$ meV ($1S$ state) and $\Delta E^{hfs}(2S)=6.2804$ meV
($2S$ state) represent reliable estimate for a comparison with forthcoming experimental data of
CREMA collaboration. The hyperfine structure interval $\Delta_{12}=8\Delta E^{hfs}(2S)-
\Delta E^{hfs}(1S)=-0.0379$ meV can be used for precision check of quantum electrodynamics predictions
for muonic deterium.
\end{abstract}



\maketitle

\section{Introduction}

In last years a considerable interest in the investigation of fine and hyperfine energy structure
of simple atoms is related with light muonic atoms: muonic hydrogen, muonic deuterium, ions of muonic
helium. This is caused by essential progress achieved by experimental collaboration CREMA
(Charge Radius Experiment with Muonic Atoms) in the study of such muonic atoms
\cite{CREMA,CREMA1}. Thus, for example, in the measurement of transition frequency
$2P^{F=2}_{3/2}-2S^{F=1}_{1/2}$ there was obtained more precise value of proton charge radius
$r_p= 0.84087(39)$ fm. The measurement of transition frequency
$2P^{F=1}_{3/2}-2S^{F=0}_{1/2}$ for singlet $2S$-state allowed to find hyperfine splitting (HFS) of
$2S$ energy level in muonic hydrogen and the value of the Zemach radius $r_Z = 1.082(37)$ fm and magnetic
radius $r_M = 0.87(6)$ fm. Analogous measurements for muonic deuterium were also completed and planned
for a publication. It is necessary to point out that the experiments of CREMA collaboration
propose one important task to improve by order of magnitude the value of charge radii in these
simple atoms (proton, deuteron, helion and $\alpha$-particle) which enter to theoretical expressions
for different fine structure intervals. For successful implementation of this program theoretical calculations
of different order corrections to fine and hyperfine structure of muonic atoms have a significant importance.
\cite{borie1,kp1996,ibk,sprung,apm,egs}. Of special note are nuclear structure corrections which can be responsible
for the solution of the proton radius puzzle \cite{CREMA,structure}.

Theoretical investigations of the energy levels of light muonic atoms were carried out
many years ago in~\cite{borie1,kp1996,ibk,sprung,egs} (other references can be found in~\cite{egs})
on the basis of the Dirac equation and nonrelativistic approach by perturbation theory in
quantum electrodynamics. An experimental activity in last years generates a need to analyze again
previous calculations in order to obtain reliably basic energy intervals: the Lamb shift
$(2P_{1/2}-2S_{1/2})$, hyperfine structure of $2S$-state, fine structure $(2P_{1/2}-2P_{3/2})$,
which could be measured in CREMA experiments in the first place. The aim of our work consists
in performing new investigation of contributions $\alpha^5$ and $\alpha^6$ in hyperfine structure
of muonic deuterium which are determined by effects of the vacuum polarization, recoil, relativistic
and deuteron structure corrections. Modern numerical values of fundamental physical constants
are taken from~\cite{MT}: the electron mass $m_e=0.510998928(11)\cdot
10^{-3}$ GeV, the muon mass $m_\mu=0.1056583715(35)$ GeV, fine structure constant
$\alpha^{-1}=137.035999074(44)$, the proton mass
$m_p$ = 0.938272046(21)~GeV, the deuteron mass $m_2=1.875612859(41)$ GeV, the deuteron magnetic moment
$\mu_d=0.8574382308(72)$ in nuclear magnetons, muon anomalous magnetic moment
$a_\mu=1.16592091(63)\cdot 10^{-3}$.

In our calculation we use the quasipotential method in quantum electrodynamics as applied to the particle
bound states~\cite{faustov}, where two-particle bound state is described by the Schr\"odinger equation.
The basic contribution to muon-deuteron interaction operator in $S$-state is determined by the Breit
Hamiltonian \cite{t4}:
\begin{equation}
\label{eq:breit}
H_B=H_0+\Delta V_B^{fs}+\Delta V_B^{hfs},~~~H_0=\frac{{\bf p}^2}{2\mu}-\frac{Z\alpha}{r},
\end{equation}
\begin{equation}
\Delta V_B^{fs}=-\frac{{\bf p}^4}{8m_1^3}-\frac{{\bf p}^4}{8m_2^3}+\frac{\pi Z\alpha}{2}
\left(\frac{1}{m_1^2}+\frac{\delta_I}{m_2^2}\right)\delta({\bf r})-\frac{Z\alpha}{2m_1m_2r}
\left({\bf p}^2+\frac{{\bf r}({\bf r}{\bf p}){\bf p}}{r^2}\right),
\end{equation}
\begin{equation}
\label{eq:bhfs}
\Delta V^{hfs}_B(r)=\frac{2\pi\alpha}{3m_1 m_p}g_dg_\mu({\bf s}_1{\bf s}_2)\delta({\bf r}),
\end{equation}
where $m_1$, $m_2$ are the muon and deuteron masses respectively, $m_p$ is the proton mass,
$g_d$, $g_\mu$ are gyromagnetic factors of the deuteron and muon. The deuteron factor $\delta_I=0$, because
we use the following definition of the deuteron charge radius $r_d^2=-6\frac{dF_C}{dQ^2}\vert_{Q^2=0}$ \cite{kms,kp1995}.
Then the basic contribution to the hyperfine splitting of $S$-wave levels (the Fermi energy) is given by spin-spin
interaction part of the potential \eqref{eq:bhfs}. Averaging \eqref{eq:bhfs} over the Coulomb wave functions of $1S$- and $2S$-states
\begin{equation}
\label{eq:1sw}
\psi_{100}(r)=\frac{W^{3/2}}{\sqrt{\pi}}e^{-Wr},~~~W=\mu Z\alpha,
\end{equation}
\begin{equation}
\label{eq:2sw}
\psi_{200}(r)=\frac{W^{3/2}}{2\sqrt{2\pi}}e^{-Wr/2}\left(1-\frac{Wr}{2}\right),
\end{equation}
we obtain the following results for the leading order contribution to hyperfine splitting:
\begin{equation}
\label{eq:hfs}
E_F(nS)=\frac{2\mu^3 \alpha^4\mu_d}{m_1m_pn^3}=
\Biggl\{{{1S:~49.0875~meV}\atop{2S:~6.1359~meV}},
\end{equation}

The Fermi energy \eqref{eq:hfs} does not contain a contribution of muon anomalous magnetic
moment (AMM). The muon AMM correction to hyperfine splitting can be presented separately
taking experimental value of muon AMM \cite{MT}:
\begin{equation}
\label{eq:amm}
\Delta E^{hfs}_{a_\mu}(nS)=a_\mu E_F(nS)=
\Biggl\{{{1S:~0.0572~meV}\atop{2S:~0.0072~meV}}.
\end{equation}
Numerical value of relativistic correction of order $\alpha^6$ to HFS can be obtained by
means of known analytical expression from~\cite{egs,breit}:
\begin{equation}
\label{eq:relat}
\Delta E^{hfs}_{rel}(nS)=\Biggl\{{{\frac{3}{2}(Z\alpha)^2E_F(1S)}\atop
{\frac{17}{8}(Z\alpha)^2E_F(2S)}}
=\Biggl\{{{1S:~0.0039~meV}\atop{2S:~0.0007~meV}}
\end{equation}

In what follows we investigate other important contributions to HFS of $S$-wave energy levels
in order to obtain reliable total result. Numerical values of different corrections are
presented for definiteness with the accuracy $10^{-4}$ meV.

\begin{figure}[htbp]
\centering
\includegraphics{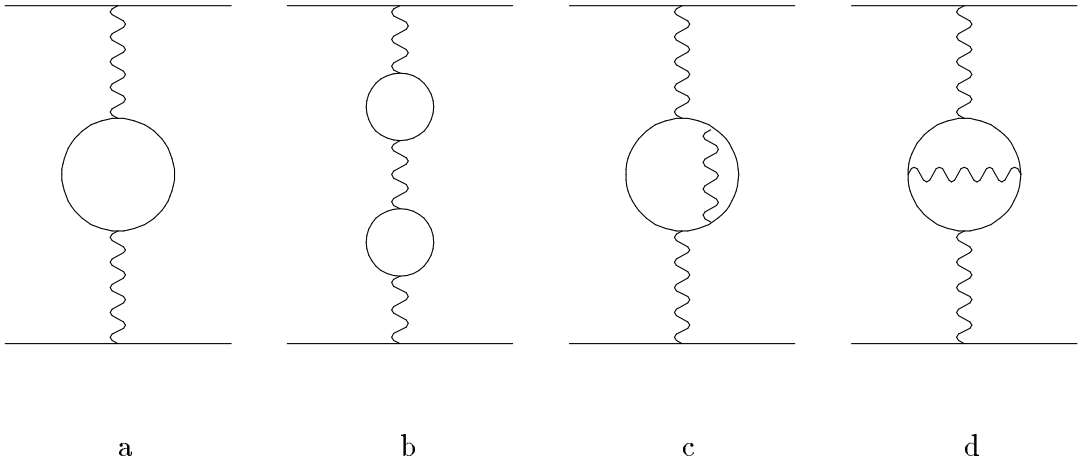}
\caption{Effects of one- and two-loop vacuum polarization in one-photon interaction.}
\label{fig:pic1}
\end{figure}

\section{Effects of one- and two-loop vacuum polarization in first order perturbation theory}

First of all, we should analyze a contribution of one-loop vacuum polarization to the potential,
which is determined in coordinate representation as follows \cite{kp1996,t4}:
\begin{equation}
\label{eq:1gammavppot}
\Delta V^{hfs}_{1\gamma,VP}(r)=\frac{4\alpha\mu_d(1+a_\mu)}{3m_1m_p}
({\bf s}_1{\bf s}_2)
\frac{\alpha}{3\pi}
\int_1^\infty\rho(s)ds\left(\pi\delta({\bf r})-\frac{m_e^2\xi^2}{r}e^{-2m_e\xi r}\right),
\end{equation}
where spectral function $\rho(\xi)=\sqrt{\xi^2-1}(2\xi^2+1)/\xi^4$. For its derivation a replacement in
photon propagator is used:
\begin{equation}
\label{eq:propcc}
\frac{1}{k^2}\to\frac{\alpha}{3\pi}\int_1^\infty\rho(\xi)d\xi\frac{1}{k^2+4m_e^2\xi^2}.
\end{equation}
We also preserve a factor with muon AMM that leads to the accounting effectively a correction of order
$\alpha^6$. Averaging~\eqref{eq:1gammavppot} over wave functions~\eqref{eq:1sw} and \eqref{eq:2sw}, we obtain the
contribution of order $\alpha^5$ to hyperfine structure of $1S-$ and $2S$-states:
\begin{equation}
\label{eq:1gammavp1s}
\Delta E_{1\gamma,VP}^{hfs}(1S)=\frac{2\mu^3 \alpha^5\mu_d(1+a_\mu)}{3m_1m_p\pi}\int_1^\infty
\rho(\xi)d\xi\left[1-\frac{m_e^2\xi^2}{W^2}\int_0^\infty xdxe^{-x\left(1+\frac{m_e\xi}{W}\right)}\right]=
0.1039~meV,
\end{equation}
\begin{equation}
\label{eq:1gammavp2s}
\Delta E_{1\gamma,VP}^{hfs}(2S)=\frac{\mu^3 \alpha^5\mu_d(1+a_\mu)}{3m_1m_p\pi}\int_1^\infty\rho(\xi)d\xi\times
\end{equation}
\begin{displaymath}
\left[1-\frac{4m_e^2\xi^2}{W^2}\int_0^\infty x\left(1-\frac{x}{2}\right)^2dxe^{-x\left(1+\frac{2m_e\xi}{W}\right)}\right]=
0.0134~meV.
\end{displaymath}

Changing the electron mass $m_e$ to muon mass $m_1$ in \eqref{eq:1gammavp1s} and \eqref{eq:1gammavp2s}, the muon vacuum
polarization contribution to HFS can be found: 0.0009 meV ($1S$), 0.0001 meV ($2S$). It has higher order
$\alpha^6$ because the ratio $W/m_1\ll 1$ and is included in Table~\ref{tb1} in corresponding line.
The same order $\alpha^6$ contribution is given also by two-loop vacuum polarization diagrams
(see Fig.~\ref{fig:pic1}(b,c,d) (the K\"allen and Sabry potential \cite{sabry}).
In order to obtain the interaction operator for the amplitude with two sequential loops
(Fig.~\ref{fig:pic1}(b)), it is necessary to use twice a replacement~\eqref{eq:propcc}. Thus in coordinate space a potential
takes a form:
\begin{equation}
\label{eq:1gamma2looppot}
\Delta V_{1\gamma,VP-VP}^{hfs}(r)=\frac{8\pi\alpha\mu_d(1+a_\mu)}{3m_1m_p}
({\bf s}_1{\bf s}_2)\left(\frac{\alpha}
{3\pi}\right)^2\int_1^\infty\rho(\xi)d\xi\int_1^\infty\rho(\eta)d\eta\times
\end{equation}
\begin{displaymath}
\times\left[\delta({\bf r})-\frac{m_e^2}{\pi r(\eta^2-\xi^2)}\left(\eta^4 e^{-2m_e\eta r}-
\xi^4 e^{-2m_e\xi r}\right)\right].
\end{displaymath}
Corresponding correction to HFS of levels $1S$ and $2S$ can be presented firstly in integral form over
coordinate $r$ and spectral parameters $\xi$ and $\eta$. After that the integration over $r$ can be done analytically
and two other integrations numerically with the use a system Mathematica. Two-loop vacuum polarization correction
of order $\alpha^6$ in Fig.~\ref{fig:pic1}(c,d) can be calculated similarly. In this case a potential of muon-deutron
interaction is determined by more complicated expression
\begin{equation}
\label{eq:1gamma2loop1pot}
\Delta V_{1\gamma,2-loop~VP}^{hfs}(r)=\frac{8\alpha^3\mu_d(1+a_\mu)}{9\pi^2 m_1m_p}({\bf s}_1{\bf s}_2)
\int_0^1\frac{f(v)dv}{1-v^2}\left[\pi\delta({\bf r})-\frac{m_e^2}{ r(1-v^2)}e^{-\frac{2m_er}{\sqrt{1-v^2}}}\right],
\end{equation}
where two-loop spectral function
\begin{equation}
f(v)=v\Bigl\{(3-v^2)(1+v^2)\left[Li_2\left(-\frac{1-v}{1+v}\right)+2Li_2
\left(\frac{1-v}{1+v}\right)+\frac{3}{2}\ln\frac{1+v}{1-v}\ln\frac{1+v}{2}-
\ln\frac{1+v}{1-v}\ln v\right]
\end{equation}
\begin{displaymath}
+\left[\frac{11}{16}(3-v^2)(1+v^2)+\frac{v^4}{4}\right]\ln\frac{1+v}{1-v}+
\left[\frac{3}{2}v(3-v^2)\ln\frac{1-v^2}{4}-2v(3-v^2)\ln v\right]+
\frac{3}{8}v(5-3v^2)\Bigr\},
\end{displaymath}
$Li_2(z)$ the Euler dilogarithm. Numerical corrections of the operator \eqref{eq:1gamma2loop1pot}
to the energy spectrum are evaluated as in the case of~\eqref{eq:1gamma2looppot}. Summary corrections
from potentials \eqref{eq:1gamma2looppot} and \eqref{eq:1gamma2loop1pot} are equal
\begin{equation}
\label{eq:hfsvpvp}
\Delta E^{hfs}_{1\gamma,VP,VP}(nS)=\Biggl\{{{1S:~0.0005~meV}\atop{2S:~0.00006~meV}},
\end{equation}
One-loop and two-loop contributions of order $\alpha^5$ and $\alpha^6$ to HFS should be considered also
in second order perturbation theory.

\begin{figure}[htbp]
\centering
\includegraphics{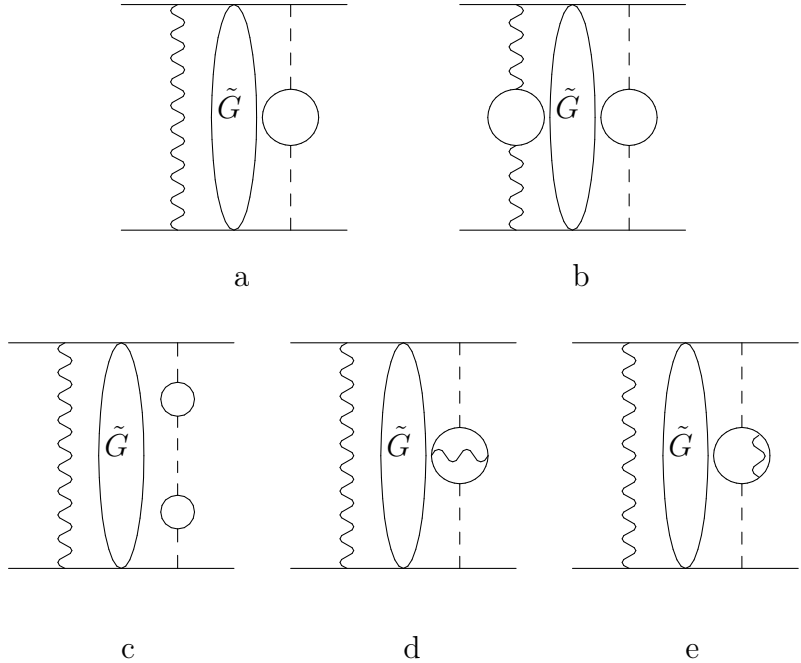}
\caption{Effects of one- and two-loop vacuum polarization in second order
perturbation theory. Dashed line denotes the Coulomb photon.
$\tilde G$ is the reduced Coulomb Green function.}
\label{fig:pic2}
\end{figure}

\section{Effects of one- and two-loop vacuum polarization in second and third order perturbation theory}

The second order perturbation theory (PT) corrections to the energy spectrum are
determined by the reduced Coulomb Green's function $\tilde G$, which has the following partial expansion:
\begin{equation}
\tilde G_n({\bf r}, {\bf r'})=\sum_{l,m}\tilde g_{nl}(r,r')Y_{lm}({\bf n})
Y_{lm}^\ast({\bf n'}).
\end{equation}
The radial function $\tilde g_{nl}(r,r')$ was obtained in~\cite{hameka} in the form of the Sturm
expansion in the Laguerre polynomials. The main contribution of the electron vacuum polarization
to HFS in second order PT has the form (see Fig.~\ref{fig:pic2}(a)):
\begin{equation}
\label{eq:soptmainf}
\Delta E^{hfs}_{SOPT~VP~1}=2<\psi|\Delta V^C_{VP}\cdot \tilde G\cdot\Delta
V_B^{hfs}|\psi>,
\end{equation}
where the modified Coulomb potential
\begin{equation}
\label{eq:modcol}
\Delta V^C_{VP}(r)=\frac{\alpha}{3\pi}\int_1^\infty\rho(\xi)d\xi\left(-\frac{Z\alpha}{r}\right)
e^{-2m_e\xi r}.
\end{equation}
Since $\Delta V_B^{hfs}(r)$ is proportional to $\delta({\bf r})$, it is necessary to use
the reduced Coulomb Green's function with one zero argument. For this case it was obtained on
the basis of the Hostler representation after a subtraction of the pole term and has the form
\cite{hameka}:
\begin{equation}
\label{eq:green1sr}
\tilde G_{1S}({\bf r},0)=\frac{Z\alpha\mu^2}{4\pi}\frac{e^{-x}}{x}g_{1S}(x),~
g_{1S}(x)=\left[4x(\ln 2x+C)+4x^2-10x-2\right],
\end{equation}
\begin{equation}
\label{eq:green2sr}
\tilde G_{2S}({\bf r},0)=-\frac{Z\alpha\mu^2}{4\pi}\frac{e^{-x/2}}{2x}g_{2S}(x),
g_{2S}(x)=\left[4x(x-2)(\ln x+C)+x^3-13x^2+6x+4\right],
\end{equation}
where $C=0.5772...$ is the Euler constant and $x=Wr$. As a result necessary corrections to HFS of
$(\mu d)$ can be presented as follows:
\begin{equation}
\label{eq:sotp1loop1s}
\Delta E^{hfs}_{VP~1}(1S)=-E_F(1S)\frac{\alpha}{3\pi}(1+a_\mu)\int_1^\infty
\rho(\xi)d\xi
\int_0^\infty e^{-x\left(1+\frac{m_e\xi}{W}\right)}g_{1S}(x)dx=0.2056~meV,
\end{equation}
\begin{equation}
\label{eq:sotp1loop2s}
\Delta E^{hfs}_{VP~1}(2S)=E_F(2S)\frac{\alpha}{3\pi}(1+a_\mu)\int_1^\infty
\rho(\xi)d\xi\int_0^\infty e^{-x\left(1+\frac{2m_e\xi}{W}\right)}g_{2S}(x)(1-\frac{x}{2})dx
=0.0207~meV.
\end{equation}
The factor $(1+a_\mu)$ is included in \eqref{eq:sotp1loop1s} and \eqref{eq:sotp1loop2s}, therefore
these expressions contain corrections of order $\alpha^5$ and $\alpha^6$. Changing $m_e\to m_1$ in
\eqref{eq:sotp1loop1s}-\eqref{eq:sotp1loop2s} we calculate one-loop muon vacuum polarization
contribution in second order PT of order $\alpha^6$: 0.0009 meV ($1S$), 0.0001 meV ($2S$). It
is included also in Table~\ref{tb1} together with similar contribution in first order PT.
Two-loop corrections in Fig.~\ref{fig:pic2}(b,c,d,e) are of order $\alpha^6$. Let us consider
first contribution which is related with potentials \eqref{eq:1gammavppot} and \eqref{eq:modcol},
reduced Coulomb Green's functions \eqref{eq:green1sr}, \eqref{eq:green2sr} and reduced Coulomb Green's
function with nonzero arguments. General structure of this contribution takes the form:
\begin{equation}
\label{eq:gensopt}
\Delta E^{hfs}_{SOPT~VP~2}=2<\psi|\Delta V^{hfs}_{1\gamma,VP}\cdot \tilde G \cdot\Delta
V^C_{VP}|\psi>.
\end{equation}
The convenient representation for reduced Coulomb Green's function with nonzero arguments was
obtained in~\cite{hameka}:
\begin{equation}
\label{eq:green1s}
\tilde G_{1S}(r,r')=-\frac{Z\alpha\mu^2}{\pi}e^{-(x_1+x_2)}g_{1S}(x_1,x_2),
\end{equation}
\begin{displaymath}
g_{1S}(x_1,x_2)=\frac{1}{2x_<}-\ln 2x_>-\ln 2x_<+Ei (2x_<)+\frac{7}{2}-2C-(x_1+x_2)+
\frac{1-e^{2x_<}}{2x_<},
\end{displaymath}
\begin{equation}
\label{eq:green2s}
\tilde G_{2S}(r,r')=-\frac{Z\alpha\mu^2}{16\pi x_1x_2}e^{-(x_1+x_2)}g_{2S}(x_1,x_2),
\end{equation}
\begin{displaymath}
g_{2S}(x_1,x_2)=8x_<-4x^2_<+8x_>+12x_<x_>-26x^2_<x_>+2x^3_<x_>-4x^2_>-
26x_<x^2_>+23x^2_<x^2_>-
\end{displaymath}
\begin{displaymath}
-x^3_<x^2_>+2x_<x^3_>-x^2_<x^3_>+4e^x(1-x_<)(x_>-2)x_>+4(x_<-2)x_<(x_>-2)x_>
\times
\end{displaymath}
\begin{displaymath}
\times[-2C+Ei(x_<)-\ln(x_<)-\ln(x_>)].
\end{displaymath}
The substitution of \eqref{eq:1gammavppot}, \eqref{eq:modcol}, \eqref{eq:green1s} and \eqref{eq:green2s} into
\eqref{eq:gensopt} provides two contributions for each $1S$ and $2S$ level in integral form:
\begin{equation}
\label{eq:27}
\Delta E^{hfs}_{VP~21}(1S)=-\frac{2\alpha^6\mu^3\mu_d(1+a_\mu)}
{9\pi^2m_1m_p}\int_1^\infty\rho(\xi)d\xi\int_1^\infty\rho(\eta)d\eta\int_0^\infty
dx e^{-x\left(1+\frac{m_e\xi}{W}\right)}g_{1S}(x),
\end{equation}
\begin{equation}
\label{eq:28}
\Delta E^{hfs}_{VP~22}(1S)=-\frac{4\alpha^6\mu^3\mu_d(1+a_\mu)m_e^2}
{9\pi^2m_1m_pW^2}\int_1^\infty\rho(\xi)d\xi\times
\end{equation}
\begin{displaymath}
\times\int_1^\infty\rho(\eta)\eta^2d\eta\int_0^\infty
x_1dx_1 e^{-x_1\left(1+\frac{m_e\xi}{W}\right)}\int_0^\infty x_2dx_2
e^{-x_2\left(1+\frac{m_e\xi}{W}\right)}g_{1S}(x_1,x_2),
\end{displaymath}
\begin{equation}
\label{eq:29}
\Delta E^{hfs}_{VP~21}(2S)=\frac{\alpha^6\mu^3\mu_d(1+a_\mu)}
{36\pi^2m_1m_p}\int_1^\infty\rho(\xi)d\xi
\int_1^\infty\rho(\eta)d\eta\int_0^\infty
\left(1-\frac{x}{2}\right)dx e^{-x\left(1+\frac{2m_e\xi}{W}\right)}g_{2S}(x),
\end{equation}
\begin{equation}
\label{eq:30}
\Delta E^{hfs}_{VP~22}(2S)=-\frac{\alpha^6\mu^3\mu_d(1+a_\mu)m_e^2}
{18\pi^2m_1m_pW^2}\int_1^\infty\rho(\xi)d\xi\times
\end{equation}
\begin{displaymath}
\times\int_1^\infty\rho(\eta)\eta^2d\eta\int_0^\infty
\left(1-\frac{x_1}{2}\right)dx_1 e^{-x_1\left(1+\frac{2m_e\xi}{W}\right)}\int_0^\infty
\left(1-\frac{x_2}{2}\right)dx_2
e^{-x_2\left(1+\frac{2m_e\xi}{W}\right)}g_{2S}(x_1,x_2).
\end{displaymath}
Separately, the contributions \eqref{eq:27}, \eqref{eq:28} and \eqref{eq:29},\eqref{eq:30}
are divergent but their sum is finite. Corresponding numerical values are presented in Table~\ref{tb1}.
The contributions of two other diagrams to HFS can be calculated by means of
\eqref{eq:soptmainf}, where the replacement of the potential \eqref{eq:modcol}
on the following potentials should be made \cite{apm}:
\begin{equation}
\label{eq:31}
\Delta V^C_{VP-VP}(r)=\left(\frac{\alpha}{3\pi}\right)^2\int_1^\infty
\rho(\xi)d\xi\int_1^\infty\rho(\eta)d\eta\left(-\frac{Z\alpha}{r}\right)
\frac{1}{\xi^2-\eta^2}\left(\xi^2 e^{-2m_e\xi r}-\eta^2 e^{-2m_e\eta r}\right),
\end{equation}
\begin{equation}
\label{eq:32}
\Delta V^C_{2-loop~VP}(r)=-\frac{2Z\alpha^3}{3\pi^2 r}\int_0^1\frac{f(v)dv}{(1-v^2)}
e^{-\frac{2m_e r}{\sqrt{1-v^2}}}.
\end{equation}
Omitting further intermediate expressions we include in Table~\ref{tb1} numerical values
of corrections from potentials \eqref{eq:29} and \eqref{eq:30}. Besides vacuum polarization
corrections there are a number of nuclear structure and recoil contributions playing essential
role in hyperfine splitting.

\begin{figure}[htbp]
\centering
\includegraphics[scale=0.6]{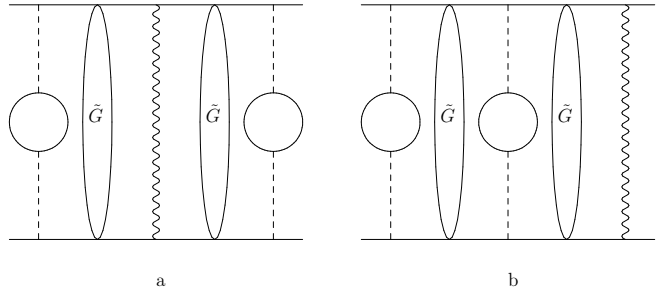}
\caption{Effects of vacuum polarization in third order
perturbation theory. Dashed line denotes the Coulomb photon.
$\tilde G$ is the reduced Coulomb Green function.}
\label{fig:pic_third}
\end{figure}

In third order perturbation theory there is a correction of order $\alpha^6$ to hyperfine splitting
which is represented symbolically in Fig.\ref{fig:pic_third} (see \cite{jetpletters2008}).
The initial expression for it is the following:
\begin{equation}
\label{eq:third}
\Delta E^{hfs}_{TOPT}=<\psi_n|\Delta V^C_{VP}\cdot \tilde G\cdot\Delta V^{hfs}\cdot\tilde G\cdot\Delta V^C_{VP}|\psi_n>+
2<\psi_n|\Delta V^C_{VP}\cdot\tilde G\cdot\Delta V^C_{VP}\cdot \tilde G\cdot\Delta V^{hfs}|\psi_n>-
\end{equation}
\begin{displaymath}
-<\psi_n|\Delta V^{hfs}|\psi_n><\psi_n|\Delta V^C_{VP}\cdot\tilde G\cdot\tilde G\cdot\Delta V^C_{VP}|\psi_n>-
\end{displaymath}
\begin{displaymath}
-2<\psi_n|\Delta V^C_{VP}|\psi_n><\psi_n|\Delta V^C_{VP}\cdot\tilde G\cdot\tilde G\cdot\Delta V^{hfs}|\psi_n>.
\end{displaymath}
The integration over one coordinate in \eqref{eq:third} is performed analytically, but second coordinate integration
and two integrations over spectral parameters are calculated numerically.
Numerically the contribution \eqref{eq:third} is essentially smaller than other corrections of order $\alpha^6$
(see Table~\ref{tb1} where it is written in separate line).

\section{Nuclear structure and recoil corrections.}

\begin{figure}[htbp]
\centering
\includegraphics[scale=0.8]{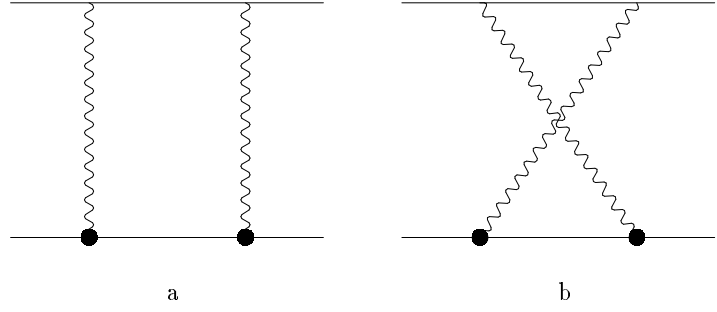}
\caption{Nuclear structure effects of order $\alpha^5$. The bold point denotes the deuteron
vertex function.}
\label{fig:pic3}
\end{figure}

The basic nuclear structure contribution to HFS of $S$-states is determined by
two-photon exchange diagrams (see Fig.~\ref{fig:pic3}). The deuteron electromagnetic current
parametrization takes the form:
\begin{equation}
J^\mu_d(p_2,q_2)=\varepsilon^\ast_\rho(q_2)\Biggl\{\frac{(p_2+q_2)_\mu}
{2m_2}g_{\rho\sigma}F_1(k^2)
-\frac{(p_2+q_2)_\mu}{2m_2}\frac{k_\rho k_\sigma}{2m_2^2}F_2(k^2)-
\Sigma^{\mu\nu}_{\rho\sigma}\frac{k^\nu}{2m_2}F_3(k^2)\Biggr\}\varepsilon_\sigma(p_2),
\end{equation}
where $p_2, q_2$ are four-momenta of the deuteron in initial and final states,
$k=q_2-p_2$. The deuteron polarization vector $\varepsilon_\mu$ satisfies to the following
conditions:
\begin{equation}
\varepsilon^\ast_\mu({\bf k},\lambda)\varepsilon^\mu({\bf k},\lambda')=-
\delta_{\lambda\lambda'},~~~ k_\mu\varepsilon^\mu({\bf k},\lambda)=0,~~~
\sum_\lambda\varepsilon^\ast_\mu({\bf k},\lambda)\varepsilon_\nu({\bf k},\lambda)=
-g_{\mu\nu}+\frac{k_\mu k_\nu}{m_2^2}.
\end{equation}
The generator of infinitesimal Lorentz transformations
is given by
\begin{equation}
\Sigma^{\mu\nu}_{\rho\sigma}=g^\mu_\rho g^\nu_\sigma-g^\mu_\sigma g^\nu_\rho.
\end{equation}
The deuteron electromagnetic form factors $F_i(k^2)$ are
functions of the square of the photon four-momentum.
They are related to the charge $F_C$, magnetic $F_M$, and quadrupole $F_Q$
form factors for the deuteron by the equations
\begin{equation}
F_C=F_1+\frac{2}{3}\eta\left[F_1+(1+\eta)F_2-F_3\right],~F_M=F_3,~F_Q=F_1+(1+\eta)F_2-F_3.
~\eta=-\frac{k^2}{4m_2^2}.
\end{equation}
The muon electromagnetic current has the form:
\begin{equation}
J_l^\mu(p_1,q_1)=\bar u(q_1)\left[\frac{(p_1+q_1)^\mu}{2m_1}-(1+a_\mu)
\sigma^{\mu\nu}\frac{k_\nu}{2m_1}\right]u(p_1),
\end{equation}
where $p_1, q_1$ are initial and final muon four-momenta,
$\sigma^{\mu\nu}$ = $(\gamma^\mu\gamma^\nu-\gamma^\nu\gamma^\mu)/2$.
The amplitudes describing the virtual Compton scattering of a muon and a deuteron
are defined by direct and crossed two-photon diagrams in the form \cite{faustov1}:
\begin{equation}
\label{eq:comp1}
M_{\mu\nu}^{(l)}=\bar u(q_1)\left[\gamma_\mu\frac{\hat p_1+\hat k+m_1}
{(p_1+k)^2-m_1^2}\gamma_\nu+\gamma_\nu\frac{\hat p_1-\hat k+m_1}
{(p_1-k)^2-m_1^2}\gamma_\mu\right]u(p_1),
\end{equation}
\begin{equation}
\label{eq:comp2}
M_{\mu\nu}^{(d)}=\varepsilon^\ast_\rho(q_2)\left[\frac{(q_2+p_2-k)_\mu}
{2m_2}g_{\rho\lambda}F_1-\frac{(q_2+p_2-k)_\mu}{2m_2}\frac{k_\rho k_\lambda}
{2m_2^2}F_2-\Sigma^{\mu\alpha}_{\rho\lambda}\frac{k_\alpha}{2m_2}F_3\right]\times
\end{equation}
\begin{displaymath}
\frac{-g_{\lambda\omega}+\frac{(p_2-k)_\lambda(p_2-k)_\omega}{m_2^2}}
{(p_2-k)^2-m_2^2}\Biggl[\frac{(p_2+q_2-k)_\nu}{2m_2}g_{\omega\sigma}F_1-
\frac{(p_2+q_2-k)_\nu}{2m_2}\frac{k_\omega
k_\sigma}{2m_2^2}F_2+
\Sigma^{\nu\beta}_{\omega\sigma}\frac{k_\beta}{2m_2}F_3\biggr]\varepsilon_\sigma(p_2).
\end{displaymath}

To construct the quasipotential of hyperfine splitting we introduce special spin projection operators
$\hat \pi_{\mu, 3/2}$ and $\hat \pi_{\mu, 1/2}$ on states with the muon-deuteron pair spin 3/2 and 1/2:
\begin{equation}
\label{eq:project}
\hat\Pi_{\mu,3/2}=[u(p_1)\epsilon_\mu(p_2)]_{3/2}=\Psi_\mu(P),~~~\hat\Pi_{\mu,1/2}=\frac{i}{\sqrt{3}}\gamma_5\left(\gamma_\mu-
v_{1,\mu}\right)\Psi(P),
\end{equation}
\begin{equation}
\label{eq:sum}
\sum_{\lambda}\Psi_\mu^\lambda(P)\bar\Psi_\nu^\lambda(P)=-\frac{\hat v_1+1}{2}
\left(g_{\mu\nu}-\frac{1}{3}\gamma_\mu\gamma_\nu-\frac{2}{3}v_{1,\mu}v_{1,\nu}+\frac{1}{3}
(v_{1,\mu}\gamma_\nu-v_{1,\nu}\gamma_\mu\right),
\end{equation}
where the spin-vector $\Psi_\mu(P)$ and spinor $\Psi(P)$ describe bound states of the muon and deuteron with spins
3/2 and 1/2, $v_{1,\mu}=P_\mu/M$, $P=p_1+p_2$, $M=m_1+m_2$. Multiplying amplitudes \eqref{eq:comp1} and \eqref{eq:comp2} and
introducing projection operators \eqref{eq:project}, we obtain by means of the package Form \cite{form} the expression for HFS
part of the potential of two-photon interaction in the Coulomb gauge for exchanged photons \cite{faustov1}:
\begin{equation}
\label{eq:v2gamma}
V_{2\gamma,str}^{hfs}=(Z\alpha)^2\int\frac{id^4k}{\pi^2}\frac{1}{(k^2)^2}
\frac{1}{k^4-4k_0^2m_1^2}\frac{1}{k^4-4k_0^2m_2^2}\times
\end{equation}
\begin{displaymath}
\left\{2F_1F_3k^6\left(\frac{k^2}{m_2^2}-\frac{{\bf k}^2}{m_2^2}-4\right)
+ 2F_2F_3\frac{k^4}{m_2^2}\left(4k_0^4+{\bf k}^4-4{\bf k}^2k_0^2-\frac{k^6}{m_2^2}\right)+
2F_3^2k^2{\bf k}^2\left(k_0^2+\frac{k^4}{m_2^2}\right)\right\}.
\end{displaymath}
The infrared divergence in~\eqref{eq:v2gamma} at $k\to 0$ is related with a term $\sim F_1F_3 k^2$.
It can be eliminated by means of iteration term of the quasipotential:
\begin{equation}
\label{eq:2gammaiter}
\Delta V_{iter}^{hfs}=\left[V_{1\gamma}\times G^f\times V_{1\gamma}\right]^{HFS}=
E_F\frac{16\mu\alpha}{3\pi n^3}({\bf S}_1{\bf S}_2)\int_0^\infty\frac{dk}{k^2}F_1F_3.
\end{equation}

\begin{table}[htbp]
\caption{Hyperfine structure of $S$-states of muonic deuterium.}
\label{tb1}
\bigskip
\begin{ruledtabular}
\begin{tabular}{|c|c|c|c|}   \hline
Contribution to hyperfine splitting  & $1S$, meV& $2S$, meV &   Reference, equation  \\   \hline
Contribution of order $\alpha^4$, the Fermi energy & 49.0875    & 6.1359&  \eqref{eq:hfs}, \cite{borie1}  \\ \hline
Muon AMM contribution  &   0.0572  & 0.0072&  \eqref{eq:amm}, \cite{borie1}  \\  \hline
Relativistic correction of order
$\alpha^6$ & 0.0039      & 0.0007 & \eqref{eq:relat}, \cite{breit},\cite{borie1}   \\  \hline
Vacuum polarization contribution of order $\alpha^5$ & 0.3095& 0.0341 &\eqref{eq:1gammavp1s}-\eqref{eq:1gammavp2s},\eqref{eq:sotp1loop1s}-\eqref{eq:sotp1loop2s},\cite{borie1}   \\     \hline
Vacuum polarization contribution of order $\alpha^6$  & 0.0048 &  0.0005&  \eqref{eq:hfsvpvp},\eqref{eq:27}-\eqref{eq:32} \\   \hline
Vacuum polarization contribution of order  & 0.0005 &  0.00004&  \eqref{eq:third}  \\
$\alpha^6$ in third order PT               &        &         &                \\   \hline
Nuclear structure correction of order $\alpha^5$&  -0.9305 &   -0.1163& \eqref{eq:2gamma_struct}   \\  \hline
Nuclear structure and vacuum polarization&  0.0152 &   0.0019& \eqref{eq:2gammastruct_vp}   \\
correction of order  $\alpha^6$ &     &      &   \\    \hline
Nuclear structure and muon vacuum polarization &  0.0015 &   0.0002& \eqref{eq:2gammastruct_mvp}   \\
correction of order  $\alpha^6$ &     &      &   \\    \hline
Hadron vacuum polarization contribution of order $\alpha^6$ &  0.0018   & 0.0002  &\eqref{eq:hvp_cor}, \cite{apm_hvp}   \\  \hline
Nuclear structure correction of order  $\alpha^6$ &  0.0082  &    0.0008&  \eqref{eq:corrm}  \\
in  $1\gamma$ interaction    &     &       &       \\   \hline
Nuclear structure correction in second &  -0.0555  &  -0.0069& \eqref{eq:1s_SOPT}-\eqref{eq:2s_SOPT}    \\
order perturbation theory    &     &      &        \\   \hline
Radiative nuclear finite size  &  -0.0039  &  -0.0005 & \eqref{eq:result1}-\eqref{eq:result3}   \\
correction of order  $\alpha^6$    &     &      &         \\   \hline
Deuteron polarizability contribution    & 1.6972   & 0.2121  &  \cite{ibk}  \\
of order $\alpha^5$ &       &       &      \\     \hline
Deuteron internal polarizability contribution    & 0.0840   & 0.0105  &  \cite{faustov1}  \\
of order $\alpha^5$ &       &       &      \\     \hline
Weak interaction contribution   &0      &0  &  \cite{egs,eides_weak}  \\   \hline
Summary contribution  &  50.2814   &  6.2804  &    \\   \hline
\end{tabular}
\end{ruledtabular}
\end{table}

\begin{figure}[htbp]
\centering
\includegraphics{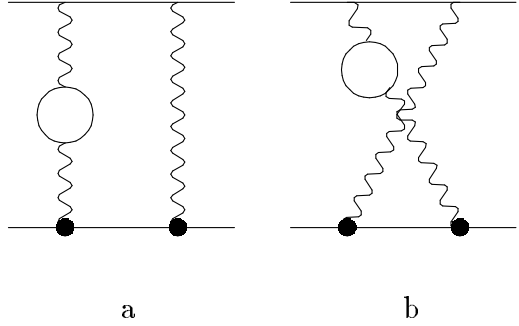}
\caption{Two photon exchange amplitudes accounting for effects of vacuum polarization and
nuclear structure. The wavy line denotes the photon. The bold point denotes the deuteron
vertex function.}
\label{fig:pic4}
\end{figure}

The angle integration in~\eqref{eq:v2gamma} in the Euclidean momentum space can be carried out analytically.
As a result the contribution of two-photon exchange amplitudes to HFS of $S$-levels can be written in the form
of one-dimensional integral:
\begin{equation}
\label{eq:2gammastruct}
E_{2\gamma}^{hfs}=E_F\alpha\int_0^\infty\frac{dk}{k^2}V_{2\gamma}(k)=
\end{equation}
\begin{displaymath}
=\frac{E_F\alpha}{16\pi m_1^3m_2^5(m_1^2-m_2^2)}
\int_0^\infty\frac{dk}{k^2}\Bigl\{4m_1^2m_2^2F_1F_3 \Bigl[k^5(m_2^2-m_1^2)+8k^2m_1^2m_2^2(h_2-h_1)+
\end{displaymath}
\begin{displaymath}
16m_1^2m_2^4(h_2-h_1)-32m_1^2m_2^4(m_2-m_1)+k^4(m_1^2h_2-m_2^2h_1)\Bigr]+
\end{displaymath}
\begin{displaymath}
+F_2 F_3k^2\Bigl[k^5(m_2^4-m_1^4)+6k^3m_1^2m_2^2(m_1^2-m_2^2)+8k^2m_1^2m_2^2\left(m_1^2(h_2-2h_1)+
m_2^2h_1\right)+
\end{displaymath}
\begin{displaymath}
16m_1^4m_2^4(h_2-h_1)+k^4(m_1^4h_2-m_2^4h_1)\Bigr]+
F_3^2k^2m_2^2 \Bigl[k^3(m_1^2-m_2^2)(5m_1^2+m_2^2)+
\end{displaymath}
\begin{displaymath}
k^2\left(-5m_1^4h_2+m_2^4h_1+4m_1^2m_2^2h_1\right)+6km_1^2m_2^2(m_1^2-m_2^2)\Bigr]\Bigr\},
\end{displaymath}
where the subtraction of iteration term of the potential \eqref{eq:2gammaiter} is performed,
$h_{1,2}=\sqrt{k^2+4m_{1,2}^2}$. Moreover, we remove a factor $F_3(0)=m_2\mu_d/Zm_p$ from the
form factor $F_3$ in~\eqref{eq:2gammastruct} as in~\eqref{eq:2gammaiter}. Numerical integration
in~\eqref{eq:2gammastruct} performed with the use of known parametrization for the deuteron form
factors \cite{abbott} gives the following result:
\begin{equation}
\label{eq:2gamma_struct}
\Delta E^{hfs}_{2\gamma,str}(nS)=\Biggl\{{{1S:~-0.9305~meV}\atop{2S:~-0.1163~meV}}.
\end{equation}
An expression \eqref{eq:2gammastruct} can be used for the evaluation of nuclear structure and vacuum
polarization corrections shown in Fig.~\ref{fig:pic4}. A change of the potential in this case as compared with
\eqref{eq:2gammastruct} can be derived after the replacement \eqref{eq:propcc} and insertion of the factor 2.
Corresponding contribution to HFS of $S$-levels is represented in the form:
\begin{equation}
\label{eq:2gammastruct_vp}
E_{2\gamma,VP}^{hfs}=-E_F\frac{2\alpha^2}{3\pi}\int_1^\infty\rho(\xi)d\xi
\int_0^\infty V_{2\gamma}(k)\frac{dk}{k^2+4m_e^2\xi^2}=\Biggl\{{{1S:~0.0152~meV}\atop{2S:~0.0019~meV}}.
\end{equation}

The muon vacuum polarization contribution is also calculated by formula~\eqref{eq:2gammastruct_vp}
after a replacement $m_e\to m_1$:
\begin{equation}
\label{eq:2gammastruct_mvp}
E_{2\gamma,MVP}^{hfs}=\Biggl\{{{1S:~0.0015~meV}\atop{2S:~0.0002~meV}}.
\end{equation}

To increase the calculation accuracy we consider hadron vacuum polarization (HVP) contribution.
A potential $V_{2\gamma}$ accounting for nuclear structure corrections is applicable for this aim.
A replacement in photon propagator for HVP correction takes the form
\begin{equation}
\label{eq:hvp}
\frac{1}{k^2}\to\left(\frac{\alpha}{\pi}\right)\int_{s_{th}}^\infty\frac{\rho_{had}(s)ds}{k^2+s},
\end{equation}
and leads to the following expression for hadron vacuum polarization correction:
\begin{equation}
\label{eq:hvp_cor}
E_{2\gamma,HVP}^{hfs}=-E_F\frac{2\alpha^2}{\pi}\int_1^\infty\rho_{had}(s)ds
\int_0^\infty V_{2\gamma}(k)\frac{dk}{k^2+s}.
\end{equation}
The basic contribution to hadron spectral function $\rho_{had}(s)$ is determined by the pion form factor
$F_\pi(s)$ for the energy interval $4m_\pi^2\div 0.81$ $GeV^2$:
\begin{equation}
\label{eq:ffpi}
\rho_{had}(s)=\frac{(s-4m_\pi^2)^{3/2}}{12s^{5/2}}|F_\pi(s)|^2.
\end{equation}
The contributions of resonances $J^{PC}=1^{--}$ of $J/\psi$ and $\Upsilon$ families and other nonresonance energy $s$
intervals are calculated as in our previous works \cite{apm_hvp}. Total hadron vacuum polarization contribution is
presented in Table~\ref{tb1}.

\begin{figure}[htbp]
\centering
\includegraphics[scale=0.8]{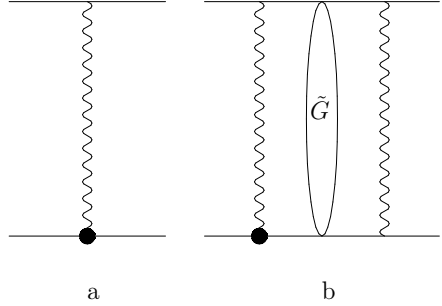}
\caption{Nuclear structure effects in one-photon interaction and in second order perturbation
theory. $\tilde G$ is the reduced Coulomb Green's function.}
\label{fig:pic5}
\end{figure}

Six order $\alpha$ corrections are also shown in Fig.~\ref{fig:pic5}. To evaluate the contribution of the amplitude in
Fig.~\ref{fig:pic5}(a) we use an expansion of deuteron magnetic form factor:
\begin{equation}
\label{eq:rm}
G_M(k^2)=\frac{m_2}{Zm_p}\mu_d\left(1-\frac{1}{6}r_M^2{\bf k}^2\right),
\end{equation}
which leads to the following potential in momentum space:
\begin{equation}
\label{eq:potrm}
\Delta V^{hfs}(k)=-\frac{4\pi\alpha\mu_d}{9m_1m_p}r_M^2({\bf s}_1{\bf s}_2){\bf k}^2,
\end{equation}
and in coordinate representation
\begin{equation}
\label{eq:potrmr}
\Delta V^{hfs}(k)=\frac{4\pi\alpha\mu_d}{9m_1m_p}r_M^2({\bf s}_1{\bf s}_2)\nabla^2\delta({\bf r}).
\end{equation}
Averaging~\eqref{eq:potrmr} over the Coulomb wave functions, we obtain an analytical expression for a correction
to HFS and its numerical values for the levels $1S$ and $2S$:
\begin{equation}
\label{eq:corrm}
\Delta E^{hfs}_{1\gamma,str}=\frac{2}{3}\mu^2\alpha^2r_M^2\frac{3n^2+1}{n^2}E_F=
\Biggl\{{{1S:~0.0082~meV}\atop{2S:~0.0008~meV}}.
\end{equation}
Another nuclear structure correction of order $\alpha^6$ to HFS of muonic deuterium is determined in second order PT by
the amplitude presented in Fig.~\ref{fig:pic5}(b). Each of the perturbation potentials in this diagram is
proportional to $\delta({\bf r})$ if we use an expansion over small transfer momentum.
As a result a contribution of second order PT is proportional to divergent expression $\tilde G(0,0)$.
To avoid an appearance of $\tilde G(0,0)$ we use the nuclear structure perturbation potential in the form:
\begin{equation}
\label{eq:poprrm}
\Delta V^{C}_{str,1\gamma}(k)=-\frac{Z\alpha}{{\bf k}^2}\left[\frac{1}{(1+\frac{k^2}{\Lambda^2})^2}-1\right]=
\frac{Z\alpha}{\Lambda^2}\frac{(2+\frac{k^2}{\Lambda^2})}{(1+\frac{k^2}{\Lambda^2})^2},~~~\Lambda=\frac{\sqrt{12}}{r_d}.
\end{equation}
A convenience of dipole parametrization for the deuteron charge form factor in this case instead of a
parametrization from~\cite{abbott} is related with a fact that in the coordinate representation we obtain under such conditions
sufficiently simple expressions:
\begin{equation}
\label{eq:pot_lam}
\Delta V^C_{str,1\gamma}(r)=\frac{Z\alpha(2+\Lambda r)}{8\pi r}e^{-\Lambda r}.
\end{equation}
Using the Green's functions~\eqref{eq:green1sr} and \eqref{eq:green2sr} the analytical integration in second order PT
can be performed. It gives the following result:
\begin{equation}
\label{eq:1s_SOPT}
\Delta E^{hfs}_{str,SOPT}(1S)=E_F(1S)\frac{\mu\alpha}{2\pi\Lambda(1+\frac{2W}{\Lambda})^4}
\Biggl\{-2\frac{W}{\Lambda}\left[4\frac{W}{\Lambda}(5+3\frac{W}{\Lambda})+13\right]-
\end{equation}
\begin{displaymath}
-16\frac{W}{\Lambda}\left(1+\frac{W}{\Lambda}\right)\left(1+\frac{2W}{\Lambda}\right)\coth^{-1}\left(1+\frac{4W}{\Lambda}\right)-3\Biggl\}
=-0.0555~meV,
\end{displaymath}
\begin{equation}
\label{eq:2s_SOPT}
\Delta E^{hfs}_{str,SOPT}(2S)=-E_F(2S)\frac{\mu\alpha}{8\pi\Lambda(1+\frac{W}{\Lambda})^6}
\Biggl\{\frac{W}{\Lambda}\left[\left(\frac{W}{\Lambda}\left(14+3\frac{W}{\Lambda}\right)+31\right)\frac{W^2}{\Lambda^2}+16\right]+
\end{equation}
\begin{displaymath}
+8\frac{W}{\Lambda}\left(1+\frac{W}{\Lambda}\right)\left[\left(3+\frac{W}{\Lambda}\right)\frac{W^2}{\Lambda^2}+4\right]\coth^{-1}\left(1+\frac{2W}{\Lambda}\right)+
6\Biggr\}=-0.0069~meV.
\end{displaymath}

\section{Radiative corrections to two photon exchange diagrams.}

\begin{figure}[htbp]
\centering
\includegraphics{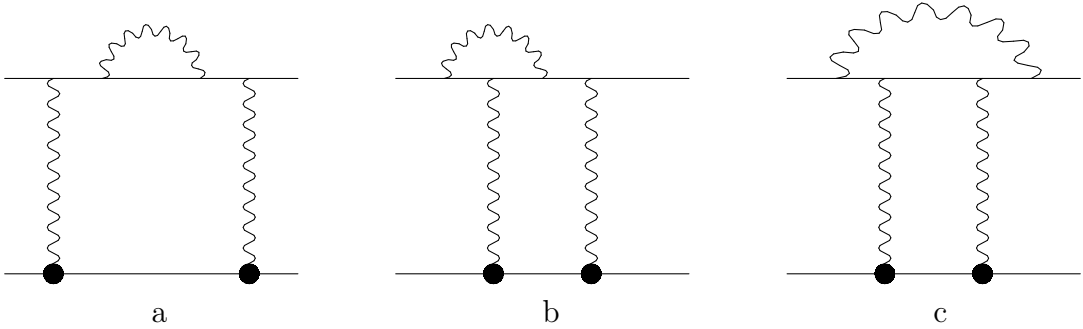}
\caption{Direct two-photon exchange amplitudes with radiative corrections to muon line
giving contributions of order $\alpha(Z\alpha)^5$ to the hyperfine structure. Wave line on the diagram denotes
the photon. Bold point on the diagram denotes the vertex operator of the proton or deuteron.}
\label{fig:pic6}
\end{figure}

The lepton line radiative corrections to two-photon exchange amplitudes are of order $\alpha(Z\alpha)^5$.
Such corrections to HFS of muonium were studied in detail in~\cite{eides1}. Total integral expression
for all radiative corrections in Fig.~\ref{fig:pic6} to HFS of order $\alpha(Z\alpha)^5$ including
recoil effects was constructed in~\cite{eides2} in Fried-Yennie gauge \cite{fried}. The advantage
of Fried-Yennie gauge in the calculation of corrections in Fig.~\ref{fig:pic6} is that it leads to
infrared finite renormalizable integral expressions for muon self-energy operator, vertex function and
lepton tensor describing the diagram with spanning photon \cite{eides3}. Using such general expressions
an analytical calculation of corrections $\alpha(Z\alpha)^5$ to HFS in the point-like nucleus approximation
can be performed. If the effects of nuclear structure should be taken into account then these expressions
allow us to obtain numerical values of diagrams in Fig.~\ref{fig:pic6}(a,b,c) separately. The muon-deuteron
scattering amplitude corresponding to direct two-photon exchange diagrams with radiative insertions in muon line
can be presented in the form:
\begin{equation}
\label{eq:2gamma}
{\cal M}=\frac{-i(Z\alpha)^2}{\pi^2}\int d^4k
\left[
\bar u(q_1)L_{\mu\nu}u(p_1)\right]D_{\mu\omega}(k)D_{\nu\lambda}(k)\times
\end{equation}
\begin{displaymath}
\left[\epsilon^\ast_\rho(q_2)\Gamma_{\omega_,\rho\beta}(q_2,p_2+k){\cal D}_{\beta\tau}(p_2+k)\Gamma_{\lambda,\tau\alpha}(p_2+k,p_2)
\epsilon_\alpha(p_2)\right]
\end{displaymath}
where $\epsilon_\rho(q)$ is the wave function of free deuteron with spin 1, $p_{1,2}$ and $q_{1,2}$ are four-momenta
of the muon and deuteron in initial and final states: $p_{1,2}\approx q_{1,2}$. The vertex operator describing the
photon-deuteron interaction is determined by three form factors as follows:
\begin{equation}
\label{eq:gamma}
\Gamma_{\omega_,\rho\sigma}(q_2,p_2+k)=\frac{(2p_2+k)_\omega}{2m_2}g_{\rho\sigma}\cdot F_1(k)-\frac{(2p_2+k)_\omega}{2m_2}
\frac{k_{\rho}k_{\sigma}}{2m_2^2}\cdot F_2(k)-(g_{\rho\gamma}g_{\sigma\omega}-
g_{\rho\omega}g_{\sigma\gamma})\frac{k_\gamma}{2m_2}\cdot F_3(k).
\end{equation}
The deuteron propagator and the photon propagator in the Coulomb gauge are equal to
\begin{equation}
\label{eq:prop}
{\cal D}_{\alpha\beta}(p)=\frac{-g_{\alpha\beta}+\frac{p_\alpha p_\beta}{m_2^2}}{(p^2-m_2^2+i0)},~~~
D_{\lambda\sigma}(k)=\frac{1}{k^2}\left[g_{\lambda\sigma}+\frac{k_\lambda k_\sigma-k_0 k_\lambda g_{\sigma 0}-
k_0 k_\sigma g_{\lambda 0}}{{\bf k}^2}\right].
\end{equation}
The lepton tensor $L_{\mu\nu}$ has a definite form for three amplitudes in Fig.~\ref{fig:pic6}. Using the package
FeynCalc \cite{fc} we perform independent construction of lepton tensors corresponding to muon self-energy, vertex
corrections and the diagram with spanning photon:
\begin{equation}
\label{eq:se}
L_{\mu\nu}^{se}=-\frac{3\alpha}{4\pi}\gamma_\mu(\hat p_1-\hat k)\gamma_\nu\int_0^1\frac{(1-x)dx}{(1-x)m_1^2+x{\bf k}^2},
\end{equation}
\begin{equation}
\label{eq:vertex}
L_{\mu\nu}^{vertex}=2\frac{\alpha}{4\pi}\int_0^1 dz\int_0^1 dx \gamma_\mu
\frac{\hat p_1-\hat k+m_1}{(p_1-k)^2-m_1^2+i0}
\left[F_\nu^{(1)}+\frac{F_\nu^{(2)}}{\Delta}+\frac{F_\nu^{(3)}}{\Delta^2}\right],
\end{equation}
\begin{equation}
\label{eq:verfun}
F_\nu^{(1)}=-6x\gamma_\nu\ln\frac{m_1^2x+{\bf k}^2z(1-xz)}{m_1^2x},~~~F_\nu^{(3)}=2x^3(1-x)\hat Q(\hat p_1-
\hat k+m_1)\gamma_\nu(\hat p_1+m_1)\hat Q,
\end{equation}
\begin{displaymath}
F_\nu^{(2)}=-x^3(2\gamma_\nu Q^2-2\hat Q\gamma_\nu \hat Q)-x^2[\gamma_\alpha\hat Q\gamma_\nu(\hat p_1+m_1)\gamma_\alpha+
\gamma_\alpha(\hat p_1-\hat k+m_1)\gamma_\nu\hat Q\gamma_\alpha+2\gamma_\nu(\hat p_1+m_1)\hat Q+
\end{displaymath}
\begin{equation}
+2\hat Q(\hat p_1-\hat k+m_1)\gamma_\nu]-x(2-x)\gamma_\alpha(\hat p_1-\hat k+m_1)\gamma_\nu(\hat p_1+m_1)\gamma_\alpha,
\end{equation}
\begin{displaymath}
Q=-p_1+kz,~\Delta=x^2m_1^2-xz(1-xz)k^2+2kp_1xz(1-x),
\end{displaymath}
\begin{equation}
\label{eq:jellyfish}
L_{\mu\nu}^{jellyfish}=\frac{\alpha}{4\pi}\int_0^1 dz\int_0^1 dx \left(\frac{F^{(1)}_{\mu\nu}}{\Delta}+
\frac{F^{(2)}_{\mu\nu}}{\Delta^2}+\frac{F^{(3)}_{\mu\nu}}{\Delta^3}\right).
\end{equation}

Explicit form of tensors $F_{\mu\nu}^{(i)}$ is presented in~\cite{fmms}. For a construction of hyperfine potential respective to the
amplitude~\eqref{eq:2gamma} we use the projection operators~\eqref{eq:project}. The insertion~\eqref{eq:project} in~\eqref{eq:2gamma}
allows us to pass to the trace calculation and a contraction over the Lorentz indices by means of the system Form~\cite{form}.
Hence, a general structure of potentials contributing to the energy shifts for states with the angular momenta $1/2$ and $3/2$ is the following:
\begin{equation}
\label{eq:pot12}
N_{1/2}=\frac{1}{6}Tr\Bigl\{\sum_{\sigma}\Psi^\sigma(P)\bar\Psi^\sigma(P)(\gamma_\rho-v_{1,\rho})\gamma_5(1+\hat v_1)
L_{\mu\nu}(1+\hat v_1)\gamma_5(\gamma_\alpha-v_{1,\alpha})\Bigr\}\times
\end{equation}
\begin{displaymath}
\Gamma_{\omega,\rho\beta}(q_2,p_2+k){\cal D}_{\beta\tau}(p_2+k)\Gamma_{\lambda,\tau\alpha}(p_2+k,p_2)D_{\mu\omega}(k)D_{\nu\lambda}(k),
\end{displaymath}
\begin{equation}
\label{eq:pot32}
N_{3/2}=\frac{1}{4}Tr\Bigl\{\sum_{\sigma}\Psi^\sigma_\alpha(P)\bar\Psi^\sigma_\rho(P)(1+\hat v_1)
L_{\mu\nu}(1+\hat v_1)\Bigr\}\times
\end{equation}
\begin{displaymath}
\Gamma_{\omega,\rho\beta}(q_2,p_2+k){\cal D}_{\beta\tau}(p_2+k)\Gamma_{\lambda,\tau\alpha}(p_2+k,p_2)
D_{\mu\omega}(k)D_{\nu\lambda}(k).
\end{displaymath}

Neglecting the recoil effects in the denominator of the deuteron propagator we obtain:
$1/[(p_2+k)^2-m_2^2+i0]\approx 1/(k^2+2kp_2+i0)\approx 1/(2k_0m_2+i0)$. The contribution of
crossed two-photon exchange diagrams is also determined by~\eqref{eq:2gamma}-\eqref{eq:jellyfish} with a
replacement $k\to -k$ in the deuteron propagator. Then a summary contribution is proportional to
$\delta(k_0)$:
\begin{equation}
\label{eq:delta}
\frac{1}{2m_2k_0+i0}+\frac{1}{-2m_2k_0+i0}=-\frac{i\pi}{m_2}\delta(k_0).
\end{equation}

As a result three types of corrections of order $\alpha(Z\alpha)^5$ to HFS of muonic deuterium
are expressed in integral form over the loop momentum ${\bf k}$ and the Feynman parameters:
\begin{equation}
\label{eq:result1}
\Delta E^{hfs}_{se}=E_F6\frac{\alpha(Z\alpha)}{\pi^2}\int_0^1 xdx\int_0^\infty\frac{F_1(k^2)F_3(k^2)dk}{x+(1-x)k^2},
\end{equation}
\begin{equation}
\label{eq:result21}
\Delta E^{hfs}_{vertex-1}=-E_F24\frac{\alpha(Z\alpha)}{\pi^2}\int_0^1 dz\int_0^1 xdx\int_0^\infty
\frac{F_1(k^2)F_3(k^2)\ln [\frac{x+k^2z(1-xz)}{x}]dk}{k^2},
\end{equation}
\begin{equation}
\label{eq:result22}
\Delta E^{hfs}_{vertex-2}=E_F8\frac{\alpha(Z\alpha)}{\pi^2}\int_0^1 dz\int_0^1 dx\int_0^\infty\frac{dk}{k^2}
\Biggl\{\frac{F_1(k^2)F_3(k^2)}{[x+k^2z(1-xz)]^2}\Bigl[-2xz^2(1-xz)k^4+
\end{equation}
\begin{displaymath}
zk^2(3x^3z-x^2(9z+1)+x(4z+7)-4)+x^2(5-x)\Bigr]-\frac{1}{2}\Biggr\},
\end{displaymath}
\begin{equation}
\label{eq:result3}
\Delta E^{hfs}_{jellyfish}=E_F4\frac{\alpha(Z\alpha)}{\pi^2}\int_0^1(1-z)dz\int_0^1(1-x)dx\int_0^\infty
\frac{F_1(k^2)F_3(k^2)dk}{[x+(1-x)k^2]^3}
\end{equation}
\begin{displaymath}
\times\Bigl[6x+6x^2-6x^2z+2x^3-12x^3z-12x^4z+k^2(-6z+18xz+4xz^2+7x^2z-30x^2z^2-
\end{displaymath}
\begin{displaymath}
2x^2z^3-36x^3z^2+12x^3z^3+24x^4z^3)+k^4(9xz^2-31x^2z^3+34x^3z^4-12x^4z^5\Bigr]
\end{displaymath}
The contribution of form factor $F_2(k^2)$ in~\eqref{eq:result1}-\eqref{eq:result3} is omitted
because the terms $F_2(k^2)F_3(k^2)$ are suppressed by powers of the mass $m_2$.
The term $1/2$ in figure brackets~\eqref{eq:result22} is related to the subtraction term of the
quasipotential. All corrections \eqref{eq:result1}, \eqref{eq:result21}, \eqref{eq:result22}, \eqref{eq:result3}
are expressed through the convergent integrals.
It is necessary to point out that expressions for the vertex function and lepton tensor with spanning photon were
obtained in~\cite{eides3} in a slightly different form. Nevertheless, in the case of point-like nucleus they lead
to contributions to hyperfine structure of $S$-states which coincide with \eqref{eq:result1}-\eqref{eq:result3}.
In numerical evaluation of finite size corrections~\eqref{eq:result1}-\eqref{eq:result3} we use the deuteron
form factor parametrization from~\cite{abbott}. Numerical results are presented in Table~\ref{tb1}.

\section{Conclusion}

In this work we calculate QED corrections, nuclear structure and recoil corrections of orders $\alpha^5$ and
$\alpha^6$ to hyperfine splitting of $1S$ and $2S$ energy levels in muonic deuterium.
In contrast to earlier performed investigations of the energy spectra of light muonic atoms in
\cite{borie1} we use three-dimensional quasipotential approach for the description of the muon-deuteron bound state.
All considered corrections due to effects of vacuum polarization and deuteron structure are
presented in integral form and calculated numerically. Numerical values of studied corrections are exhibited
in Table~\ref{tb1}. We present in Table~\ref{tb1} relevant references on equations which allow to analyze again the
sources of corresponding corrections. In line 5 of Table~\ref{tb1} we present a sum of corrections of order $\alpha^6$
which includes two-loop electron vacuum polarization corrections in first and second orders of perturbation
theory and muon one-loop vacuum polarization corrections in first and second order of perturbation theory.

As pointed out above the hyperfine structure of muonic atoms was investigated in \cite{borie1,kp1996}.
In these papers the transition frequencies between the energy levels $2P$ and $2S$ were obtained
in the case of muonic hydrogen, muonic deuterium and ions of muonic helium. The only detailed calculation
of $2S$-state hyperfine splitting in muonic deuterium is presented in~\cite{borie1}.
The splitting formula obtained in this paper
\begin{equation}
\label{eq:borie}
\Delta E_{2s} =\frac{3}{2}\beta_D(1+a_\mu)(1+\epsilon_{VP}+\epsilon_{vertex}+\epsilon_{Breit}+\epsilon_{Zemach})= 6.0584(7)~meV'
\end{equation}
contains basic corrections to the Fermi energy: the vacuum polarization, relativistic correction, vertex correction and the
Zemach contribution. Note that the Zemach correction (-0.1177(7)) meV for $2S$-state in muonic deuterium from \cite{borie1} is slightly different
from our value (-0.1163) meV what may be related with recoil effects. One-loop electron vacuum polarization corrections
in first order $\epsilon_{VP1}=0.00218$ and in second order PT $\epsilon_{VP2}=0.00337$ for $2S$-state from \cite{borie1} coincide exactly with
our results expressed by \eqref{eq:1gammavp2s} and \eqref{eq:sotp1loop2s}.
As it follows from our Table~\ref{tb1} the total value of $2S$-state hyperfine splitting 6.0683 meV without an account of deuteron
polarizability contribution obtained in \cite{ibk}
is in good agreement with the result 6.0584 meV' obtained in~\cite{borie1}. A small difference in the results is conditioned
first of all by nuclear recoil, structure and polarizability effects in two-photon exchange diagrams
which we calculate using modern experimental data on deuteron electromagnetic form factors, and by nuclear structure corrections
of order $\alpha^6$. We include in Table~\ref{tb1} numerical value of deuteron polarizability contribution to hyperfine splitting in muonic
deuterium evaluated by means of analytical expression obtained in \cite{ibk} for electronic deuterium
in the zero range approximation. This is main part of polarizability correction.
The estimate of internal deuteron polarizability correction is made in Table~\ref{tb1} on the basis of results for
muonic hydrogen. This contribution is now under consideration as in our work~\cite{faustov1}.

Assuming that the deuteron electromagnetic form factor parameterizations were obtained with an uncertainty near
0.5 per cent at small values of photon momentum squared $Q^2$ we obtain that
theoretical error in the calculation of basic nuclear structure contribution of order $(Z\alpha)^5$
which is determined by a product of two electromagnetic form factors can not be less than one per cent or $\pm 0.0010$
meV for $2S$-state. Another source of the error is related with recoil effects of order
$\alpha(Z\alpha)^5m_1/m_2$ in amplitudes in Fig.~\ref{fig:pic6}, which can amount the value 0.00002 meV.
The error in determination of internal polarizability correction is estimated approximately on muonic hydrogen
in 0.0025 meV ($2S$-state) (near 25 per cent). The estimate of an uncertainty in deuteron polarizability correction
is given on the basis of results from \cite{ibk}. It is equal approximately to 0.0042 meV (2 per cent).
Weak interaction
contribution is equal to zero in nonrelativistic approximation as was demonstrated in~\cite{egs,eides_weak}.
Our total theoretical error is estimated in 0.0050 meV in the case of $2S$-state. To obtain this estimate we add
the above mentioned uncertainties in quadrature.
The hyperfine structure interval $\Delta_{12}$ does not contain uncertainties with nuclear structure and polarizability.
So, the obtained in this work value $\Delta_{12}=-0.0379$ meV' can be used for additional check of quantum electrodynamics
in the case of muonic deuterium with a precision 0.001 meV.
To construct the quasipotential corresponding to
amplitudes in Fig.~\ref{fig:pic3} we develop the method of projection operators on the bound states with
definite spins. It allows to employ different systems of analytical calculations
\cite{form,fc}. In this approach more complicated corrections, for example, radiative recoil corrections
to hyperfine structure of order $\alpha(Z\alpha)^5m_1/m_2$ can be evaluated if an increase of the accuracy
will be needed. The results from Table~\ref{tb1} should be taken into account
for a comparison with experimental data \cite{CREMA}.

We are grateful to F.~Kottmann and E.~Borie for valuable information about CREMA experiments,
critical remarks and useful discussion of different questions related with the energy levels of light muonic atoms.
The work is supported by the Russian Foundation for Basic Research (grant 14-02-00173)
and Dynasty Foundation.

\end{document}